\documentclass[accepted]{aastex631}
\usepackage{graphicx}
\usepackage{txfonts}
\usepackage{amsfonts}
\usepackage{amssymb}
\usepackage{xcolor}
\usepackage{float}
\definecolor{seagreen}{rgb}{0.190, 0.525, 0.361}
\definecolor{ceruleanCrayola}{rgb}{0.224, 0.663, 0.859}
\definecolor{unberableRed}{rgb}{0.8, 0.06, 0.03}
\definecolor{anthracite}{rgb}{0.271, 0.270, 0.318}
\definecolor{refereebrown}{rgb}{0.28, 0.01, 0.01}
\definecolor{operamauve}{rgb}{0.718 0.518 0.655}
\definecolor{amber}{rgb}{1.0, 0.49, 0.0}
\definecolor{midnightblue}{rgb}{0.052, 0.126, 0.158}

\newcommand{\REFEREE}[1]{#1}

\shorttitle{Binary or RR Lyrae?}
\shortauthors{Trevisan et al.}
\graphicspath{{./}{figures/}}
\begin{document}
\title{Sparse logistic regression for \REFEREE{RR Lyrae vs binaries classification}}
\author[0000-0001-9511-4649]{Piero Trevisan} 
\affil{Department  of  Physics,  Università di Roma Sapienza, Piazzale Aldo Moro 2, 00185 Rome, Italy}
\affil{Ciela, Institute for Computation and Astrophysical Data Analysis , Montreal, Quebec, Canada}
\affil{INAF-Osservatorio Astronomico di Roma, via Frascati 33, 00040 Monte Porzio Catone, Italy}
\author[0000-0003-3784-5245]{Mario Pasquato}
\affiliation{Physics and Astronomy Department Galileo Galilei, University of Padova, Vicolo dell’Osservatorio 3, I–35122, Padova\\}

\affiliation{Département de Physique, Université de Montréal, Montreal, Quebec H3T 1J4, Canada\\}
\affiliation{Mila - Quebec Artificial Intelligence Institute, Montreal, Quebec, Canada}
\affil{Ciela, Institute for Computation and Astrophysical Data Analysis , Montreal, Quebec, Canada}
\author[0000-0003-3427-183X]{Gaia Carenini}
\affiliation{Department of Computer Science,
   ENS, PSL Research University,
   Paris, France\\}
\affiliation{Scuola Universitaria Superiore IUSS Pavia, Palazzo del Broletto, piazza della Vittoria 15, I-27100 Pavia, Italy\\}
\author[0000-0002-6300-6018]{Nicolas Mekhaël}
\affil{Département de Physique, Université de Montréal, Montreal, Quebec H3T 1J4, Canada}
\author[0000-0001-7511-2830]{Vittorio F. Braga}
\affil{INAF-Osservatorio Astronomico di Roma, via Frascati 33, 00040 Monte Porzio Catone, Italy}
\affil{Space Science Data Center, via del Politecnico snc, 00133 Roma, Italy}
\author[0000-0002-4896-8841]{Giuseppe Bono}
\affil{Department  of  Physics,  Universit\`a  di  Roma  Tor  Vergata,  via della Ricerca Scientifica 1, 00133 Roma, Italy}
\affil{INAF-Osservatorio Astronomico di Roma, via Frascati 33, 00040 Monte Porzio Catone, Italy}
\author[0000-0001-9146-0421]{Mohamad Abbas} 
\affiliation{Center for Astro, Particle and Planetary Physics (CAP$^3$), New York University Abu Dhabi\\}
\affil{New York University Abu Dhabi, Physic Department, PO Box 129188, Abu Dhabi, United Arab Emirates}
\begin{abstract}
\noindent 
RR Lyrae (RRL) are old, low-mass  \REFEREE{radially pulsating variable stars} in their core
helium burning phase. They are popular stellar tracers and primary
distance indicators, since they obey to \REFEREE{well} defined period-luminosity
relations in the near-infrared regime. Their photometric identification  is not trivial, indeed,
RRL samples can be contaminated by eclipsing binaries, especially in
large datasets produced by fully automatic pipelines. Interpretable machine-learning approaches for separating eclipsing binaries from RRL are thus needed. Ideally, they should be able to achieve high precision in identifying RRL while generalizing to new data from different instruments. In this paper, we train a simple logistic regression classifier on \emph{Catalina Sky Survey (CSS)} light curves. It achieves a precision of $87\%$ at $78\%$ recall for the RRL class on unseen CSS light curves. It generalizes \REFEREE{on} out-of-sample \REFEREE{data} (ASAS/ASAS-SN light curves) with a precision of \REFEREE{$85\%$} at \REFEREE{$96\%$} recall. We also considered a L1-regularized version of our classifier, which reaches $90\%$ sparsity in the light-curve features with a limited trade-off in accuracy on our CSS validation set and -remarkably- also on the ASAS/ASAS-SN light curve test set. Logistic regression is natively interpretable, and regularization allows us to point out the parts of the light curves that matter the most in classification. We thus achieved both good generalization and full interpretability.
\end{abstract}
\accepted{24 April 2023}
\section{Introduction} \label{sec:intro}


\noindent RR Lyrae (RRL) are old ($t>10$ Gyr), low-mass ($M = 0.6-0.8 M_\odot$) variable stars. \REFEREE{They typically pulsate in two main radial modes: fundamental and first overtone. A small fraction ($5$-$10\%$) pulsates simultaneously in both modes. The reader interested in a more detailed discussion concerning the possible occurrence of non-radial modes in RRL is referred to \citet{2005AcA....55...59S}.
RRL obey to \REFEREE{tight} Period-Luminosity (PL) relations for wavelengths longer than the \REFEREE{red end of the visible spectrum  \citep[][]{2004ApJS..154..633C}}. Moreover, the slope of the PL relation becomes steadily steeper and the standard dispersion steadily decreases when moving from shorter to longer wavelenghts. Near infrared PL relations provide individual distances 
with an \REFEREE{outstanding} accuracy of $\approx$ 3\% \citep{bono03c,braga15,beaton2018}.} 
\REFEREE{Hence,} they are a fundamental stepping stone in the calibration of secondary distance indicators, in turn constraining possible systematics in the cosmic distance scale \citep{beaton2016}.
\noindent Compared to other distance indicators and stellar tracers, RRL have several \REFEREE{advantages: they are ubiquitous, having been identified} both in gas-poor and in gas-rich stellar systems; \REFEREE{their lifetimes of the order of 10$^8$ yrs ensure that they outnumber other evolved radial variables
such as type II cepheids, anomalous cepheids, and classical cepheids (Bono, Braga, Pietrinferni, ARA\&A submitted); they cover a very broad metallicity distribution ranging from [Fe/H]$\sim$-3.5 to super-solar iron abundance 
\citep{crestani2021a,crestani2021b,fabrizio2021}, so they can be easily identified not only in the MW halo, but also in metal-rich stellar systems such as the MW Disk.}   \\


\noindent \REFEREE{The main issue with relying on RRL to accurately measure distances is due to their periods and amplitudes partially overlapping with eclipsing binaries, leading to spurious contamination in RRL samples. In the period-amplitude plane, also called Bailey diagram, the location of first overtone RRL overlaps with $\delta$ Scuti variables and eclipsing contact binaries. Furthermore the so-called W Ursae Majoris (W UMa) variables have sinusoidal light curves similar to RRc variables, leading to overlap in diagnostics based on Fourier parameters, especially in cases where the sampling of the light curve is either poor or the photometry is noisy so that the typical secondary eclipse of W UMa light curve can be barely detected. While e}clipsing binaries can have similar periods and similar light curve shapes to RR Lyrae, they obviously do not follow the same period-luminosity relation, thus degrading the quality of the measured distances. This is the main reason why RRc variables are only partially used in distance and in metallicity estimates 
\citep{contreras2018}.\\ 
\noindent\REFEREE{Consequently, accurate classification of RRL versus eclipsing binary stars has strong implications for our ability to avoid biases in distance determination. This has long been a priority in catalog building efforts, as in the work by \cite{2014ApJS..213....9D}, who used a robust version of the skewness of the magnitude distribution introduced by \cite{2006AJ....132.1202K} to separate the two classes in Catalina Sky Survey (CSS) data. It is thus natural to employ machine learning methods to address this task.} Since the large-scale diffusion of deep learning, more and more classification tasks in astrophysics have been entrusted to \REFEREE{various neural network architectures} \citep[see e.g.][]{2018AJ....156....7H, 2018MNRAS.473.3895L, 2019MNRAS.484..282G, 2019A&A...621A.103B, 2020MNRAS.498.5798T, 2021MNRAS.504.5656C, 2022arXiv220612402L}. Unfortunately, the excellent performance of such tools is \REFEREE{often} accompanied by insufficient theoretical guarantees \REFEREE{as well as a lack of human interpretability}. Consequently, making use of deep networks for extracting high-reliability patterns to be interpreted as physical models is challenging. While physics-informed deep learning \citep[see e.g.][]{2019JCoPh.378..686R, 2020Sci...367.1026R, 2021CMAME.373k3552H} holds the promise to mitigate this issue, in this paper we choose to favor a statistical foundation of the classifier that ensures its reliability. Moreover, in the context of scientific research and astronomy in particular, the ability to interpret our classifier is crucial. \REFEREE{It has recently been convincingly argued that n}atively interpretable models are thus preferable to black boxes even when the latter are explained a posteriori using relevant techniques \REFEREE{because explanations cannot possibly be faithful everywhere or they could simply be used in place of the underlying black box} \citep[][]{rudin2019stop}.\\

\noindent\REFEREE{Several works have already appeared in the literature attempting to address the problem of automatic classification of variable stars, but in general the main priorities of these were different from interpretability. In a recent investigation, \cite{2022MNRAS.517.3660P} computed a wide range of features \citep[including the Feature Analysis for Time Series (FATS features) introduced by][]{2015arXiv150600010N} on light curve data and apply dimensionality reduction and clustering. They also leverage the results to perform semi-supervised learning achieving a precision of $90$\% for main variable classes with a support-vector machine, thus reducing the need for human assigned labels. On the other hand,\cite{2018AJ....155...16C}  focussed their attention on modelling errors, so they apply Gaussian process regression to light curves -as we do in the following- and use it as a basis for bootstrapping and calculate FATS features, using them to train a random forest. The goal of a variety of works employing a range of neural network architectures is arguably speed and accuracy of classification, e.g. \cite{2020MNRAS.493.2981B}, \cite{2022ApJ...938...37S}, \cite{2021MNRAS.505..515Z}, \cite{2020MNRAS.493.2981B}, \cite{2019MNRAS.482.5078A}, and \cite{2007A&A...475.1159D}. }


\vspace{0.2cm}\noindent The paper is structured as follows. In Section \ref{sec:data}, we discuss the dataset, point out some of its properties obtained by a preliminary data analysis and describe the subsample involved in the experiments; afterwards, in Section \ref{giaigiaia}, we present the statistical methods involved in our classification protocol and provide justification concerning the implementation choices; subsequently in Section \ref{sec:analysis}, we give an overview of the data analysis by specifying the input representation, the validation scheme and the performance measures; then in Section \ref{sec:res}, we present the experimental results that are discussed further in Section \ref{sec:conclusion} where we describe, as well, the possible directions for further work.

\section{Data}
\label{sec:data}
\subsection{Observations}
\noindent The \REFEREE{training} data for this study comes from the Catalina Sky Survey \citep[CSS; ][]{drake2013a}. A brief summary of the characteristics of the CSS is given in \cite{2022ApJ...930..161P}, which shares part of the light curve sample with this study. A catalog of variable stars observed by CSS was published in a series of papers (e.g., \citealt{drake2013a,drake2013b,drake2014,torrealba2015,drake2017}).
Variable sources were identified based on the Welch-Stetson variability index ($I_{WS}$; \citealt{welch1993}). We initially adopt the nominal period found by the authors through Analysis of Variance (AoV; \citealt{schwarzenberg1989}); however we discuss below how this may implicitly encode information on the assumed classification of a given variable star, and the consequent pre-processing decisions we took.
The combined CSS variable catalog contains $\sim$ 110,000 variable sources, each of which was observed on average at $\sim$ 200 epochs. As pointed out also in \cite{2022ApJ...930..161P}, CSS does not cover the sky in the proximity of the Galactic plane, which in principle may introduce a bias in our results. Our choice of CSS is however justified by its large number of relatively homogeneous light curves, which is a crucial prerequisite for training most ML models.\vspace{0.2cm}\\
\REFEREE{Furthermore, to improve the quality of our variable star sample and remove any potential false positives, we undertook several steps. One of these steps involved utilizing the RUWE values obtained from Gaia \citep[][]{brown2016gaia}. To do this, we selected a circle with a radius of 3 arcsec, centered at the positions provided by CSS, and cross-matched it with the Gaia DR3 database  \citep[]{vallenari2022gaia}. We then removed all stars with RUWE values greater than 1.4. This value was chosen as a threshold, as RUWE values larger than 1.4 are indicative of potential eclipsing binaries, which are not single stars and therefore not suitable for our RRc sample. By implementing this methodology, we were able to eliminate potential false positives and increase the efficiency of our variable star sample.}
\REFEREE{We point out that CSS images were collected without filters. The conversion to the V-band was performed by setting up a sample of comparison stars (G0-G8 dwarfs) from the  2MASS Point Source Catalog \citep{skrutskie2006}, converting to the $JHKs$ bands (first step) and then to the $V$ band (second step) \citep[see][for details]{larson2003, skrutskie2006}}.
\begin{table}[H]
    \centering
    \caption{Adopted data set summary. The initial number of stars (second column) is shown broken down by variable type (first column). The third column shows the number of stars with a good Gaussian process fit according to visual inspection; only these stars were included in the final adopted data set. Median period and amplitude for each variable type are shown as well.}
    \begin{tabular}{lllll}
    \hline
    \hline
         Variable Type & N & $N_{gf}$ & Median Period & Median Amplitude \\
        \hline
         RRc & 1429 & 1298 & 0.33 & 0.40\\
         EW & 8803 & 7994 & 0.35 &0.40 \\
         EA & 2509 & 2446 & 1.03 & 0.47\\
\hline
    \end{tabular}
    
    \label{funnel}
\end{table}

\begin{figure}
    \centering
    \includegraphics[width=0.7\textwidth]{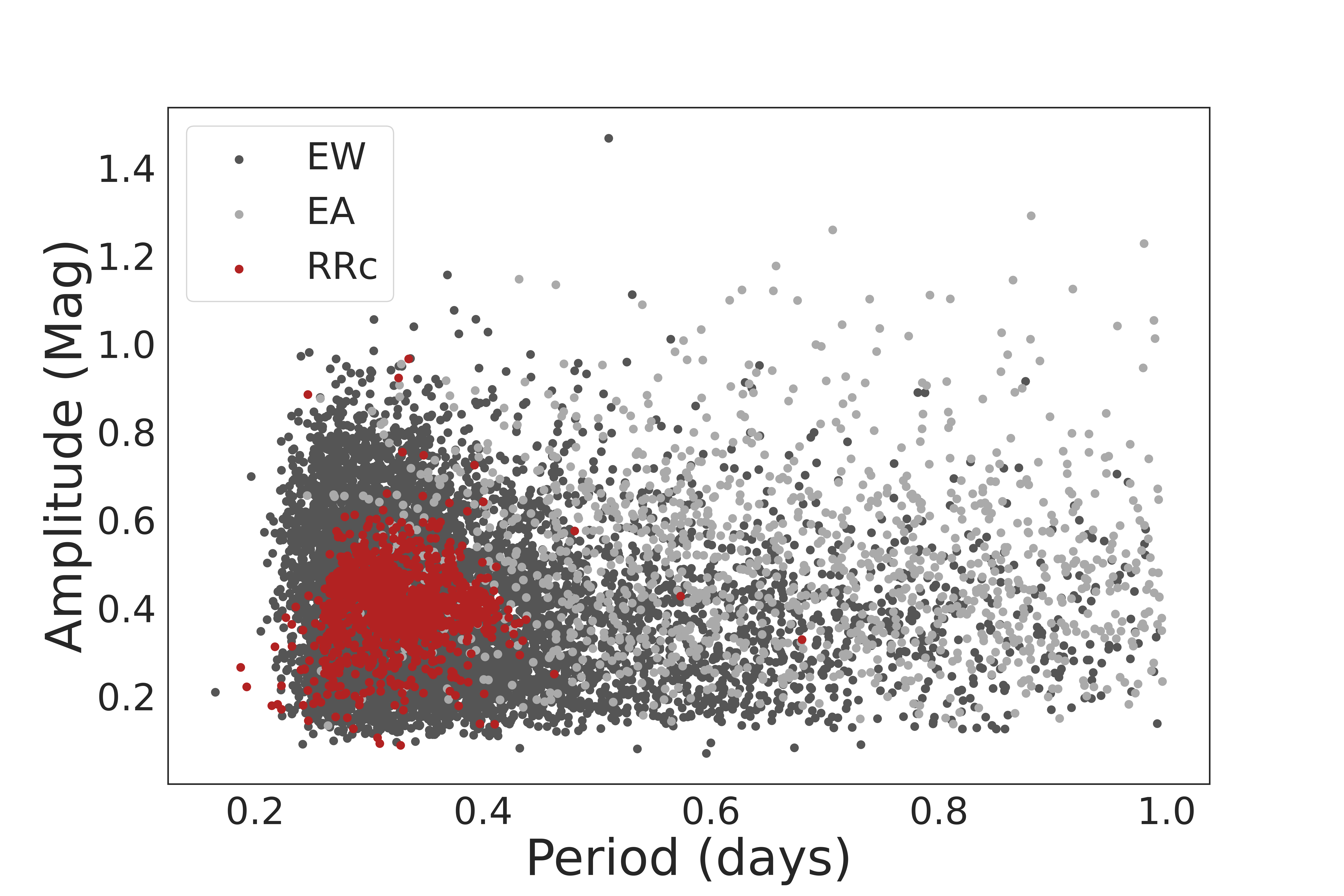}
    \caption{Period-amplitude plot of our dataset. Only the region up to 1.5 mag in amplitude and 1 day in period is shown. The stars labeled as RRc are shown in red, the eclipsing variables are shown in different shades of grey. Substantial overlap between the two groups is readily evident.}
    \label{fig:bailey_diagram}
\end{figure}

\noindent We ran our analysis on the subsample of light curves described in Tab.~\ref{funnel}; this includes both intrinsically variable RR Lyrae and eclipsing binaries of two subclasses: Algol type stars (EA) and W Uma type (EW). Even though in the initial Catalina data set the light curves of RRab stars were available, we did not include them in our analysis (except for showing a sample light curve in Fig.~\ref{fig:gaussian fit}) because they are easier to distinguish from eclipsing binary interlopers due to their characteristic shape. The Bailey diagram for the full sample is shown in Fig.~\ref{fig:bailey_diagram}. It is clear from the figure that achieving a separation between RRc and eclipsing variables is impossible solely based on period and amplitude information. Typical light curves for different variable types are shown as red points with error bars in Fig.~\ref{fig:gaussian fit}. The superimposed shaded areas are obtained by fitting the curves with a Gaussian process, as described in Sect.~\ref{giaigiaia}.\vspace{0.2cm}\\ 
\noindent To test the performance of our classifier outside of the sample it was trained on, we relied on a test set comprised of $614$ stars from the All Sky Automated Survey data \citep[ASAS; ][]{pojmanski1997, pojmanski2002} and  All-Sky Automated Survey for Supernovae \citep[ASAS-SN;][]{shappee2014,jayasinghe2019}.  This dataset differs from our CSS dataset because it was taken by a different instrument and also displays a different class balance: \REFEREE{RRL are $12\%$ of our CSS sample and $35\%$ of our ASAS/ASAS-SN test sample}. 

\subsection{Inclusion criteria}
\noindent We selected RRc, EW and EA stars only, because RRab stars are less likely to be contaminated by misclassified binary stars given the distinctive shape of their light curves. We \REFEREE{visualize our final adopted sample} in the space of the first few principal components and \REFEREE{directly inspected} the light curves \REFEREE{of any outliers}. The limited number of outliers corresponds to light curves where the Gaussian process interpolation (see below) failed to give periodic results \REFEREE{and this was evident upon visual inspection}. We speculate that this may be related to underestimated error bars in the raw data. The final adopted sample contains $11738$ stars and is described in Tab.~\ref{funnel}.

\newpage
\section{Pre-processing and classification algorithm}
\label{giaigiaia}

\subsection{Kernelized Gaussian Process}
\noindent Gaussian processes have proved to be a very versatile mathematical tool for interpolation and are progressively becoming more widespread in astronomy. Formally, a Gaussian process $\mathcal{GC}$ is a distribution over functions $f(\textbf{x})$:
\begin{equation}
f(\textbf{x})\sim \mathcal{GC}(m(\textbf{x}),k(x_i,x_j))   
\end{equation}
where $\textbf{x}$ is the \emph{function value}, $(x_i,x_j)$ are all possible pair in the input domain, $m(\textbf{x})$ is the \emph{mean function} and $k(x_i,x_j)$ is a \emph{positive definite covariance function} also known as the \emph{kernel} function.\\
For any finite subset $\mathcal{X}=\{x_1,\dots,x_n\}$ of the domain of $\textbf{x}$, the marginal distribution is a multivariate Gaussian distribution: 
\begin{equation}
f(\mathcal{X})\sim\mathcal{N}(m(\mathcal{X}),k(\mathcal{X},\mathcal{X}))   
\end{equation}
with mean vector $\mu=m(\mathcal{X})$ and covariance matrix $\Sigma=k(\mathcal{X},\mathcal{X})$. Roughly speaking, each input to this function is a variable correlated with the other variables in the input domain, as defined by the covariance function. Since functions can have an infinite input domain, the Gaussian process can be interpreted as an infinite dimensional Gaussian random variable.   \vspace{0.2cm}\\
Sampling functions from a Gaussian process boils down to fixing the mean and covariance functions. In particular, the covariance function $k(x_i,x_j)$ models the joint variability of the Gaussian process random variables and returns the modelled covariance between each pair in $x_i$ and $x_j$. Its specification force properties of the distribution over functions $f(x)$, i.e. by choosing a specific kernel function, it is possible to set prior information on this distribution. For the purpose of this work, we use a combination of four distinct kernels. The first of them belongs to the family of \emph{matern kernels} and forces the smoothness of the resulting function allowing for better interpolation. This kernel is given by: 
\begin{equation}
k(x_i,x_j)=\frac{1}{\Gamma(\nu)2^{\nu-1}}\left(\frac{\sqrt{2v}}{l}d(x_i,x_j)\right)^{\nu}K_{\nu}\left(\frac{\sqrt{2v}}{l}d(x_i,x_j)\right)
\end{equation}
where $\nu$ is a parameter controlling the smoothness of the resulting function,  $d(.,.)$ is the Euclidean distance, $K_{\nu}(.)$ is a modified Bessel function and $\Gamma(.)$ is the gamma function. The second is a \emph{white kernel}  explaining the noise of the signal as independently and identically normally-distributed. The third one is the one described by the following equation: 
\begin{equation}
    k(x_i,x_j)=\exp\left(-\frac{
    2\sin^2(\pi d(x_i,x_j)/p\REFEREE{)}}{\ell^2}\right)
\end{equation}
where $\ell$ is the length scale of the kernel, $p$ its periodicity and $d(.,.)$ the Euclidean distance. This kernel is commonly used to model functions which repeat themselves exactly. Last, we also added a \emph{constant kernel} to establish the mean of the Gaussian process.

\begin{figure}
    \centering
    \includegraphics[width=0.8\textwidth]{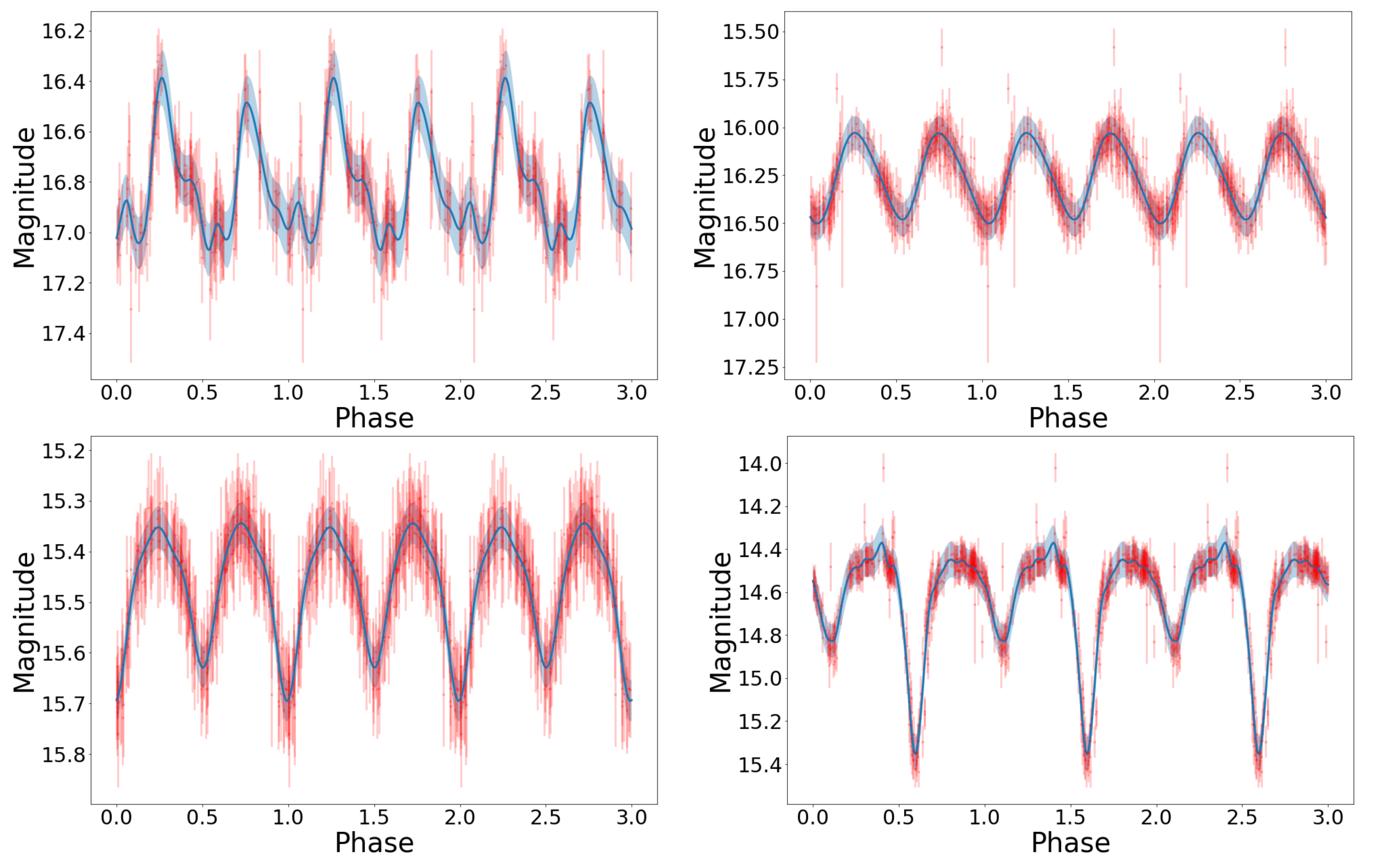}
    \caption{Example \REFEREE{phase-folded light curves} of different variable types, shown as red points with error bars: RRab (top left panel), RRc (top right panel), EW (bottom left panel) and EA (bottom right panel). The curves are repeated for three periods and were fit with a Gaussian process following the procedure described in Sect.~\ref{giaigiaia}. The solid blue line is the Gaussian process fit with the corresponding variance (light blue area).}
    \label{fig:gaussian fit}
\end{figure}

\paragraph{Application}
To obtain equally spaced data by interpolation, we applied the above specified \REFEREE{Gaussian Process to the phase folded light curves, which was phased according to} the nominal period from CSS for EA and EW variables, and to double the nominal period for the RRc variables. This allows us to compare the two on an equal footing, \REFEREE{because a miscalculated period may result in mis-classification. In particular an eclipsing binary folded to half its actual period will be much more likely to be classified as an intrinsic variable and vice-versa}. We repeated the resulting phased light curve three times to enforce periodicity. \REFEREE{
The kernel function selected for this study included the exponential component with a length scale $l=2$ and a periodicity $p=3$, combined with the \emph{constant kernel} and a \emph{white kernel} components with noise levels of 5000. We also included the \emph{matern kernel}  with a length scale $l=5$ and a $\nu=3/2$. It \REFEREE{is} worth noting that the kernel parameters were kept constant throughout the study, and no optimization was performed. This means that only one kernel function with fixed parameters was used for all stars. 
} 
\noindent We used the Python library scikit-learn \citep[][]{scikit-learn} based on  \cite{rasmussen2003gaussian} to fit our Gaussian process model. 
The resulting interpolated light curve is sampled at $300$ equally spaced points in phase, normalized between $0$ and $1$.   \REFEREE{Finally, we  phase-shifted to make the maximum coincide with the beginning of the period and then, we selected 100 consequent points. By operating in this way we avoided any possible edge effect that could affect our final dataset.}
\REFEREE{Not only}, we want our classifier to rely only on the light curve topology, we select one period in order to have all the data in a box $[0,1]\times[0,1]$. Thus each curve is represented by $100$ features corresponding to normalized values of the magnitude at given phases. The results of the Gaussian process fit are shown in Fig.~\ref{fig:gaussian fit}.

\begin{figure}
    \centering
    \includegraphics[width = .5\linewidth]{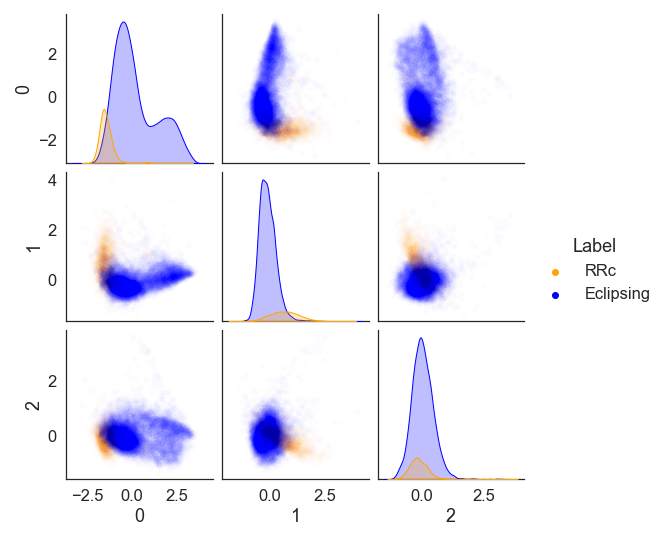}
    \caption{Distribution of our data in principal component space. The first three components are shown. The RRc variables are shown in orange, the eclipsing binaries in light blue.}
    \label{hukkusopalopem}
\end{figure}

\vspace{0.2cm}\noindent For purposes of visualization, we calculated the first three principal components of our data set in feature space. These are shown in Fig.~\ref{hukkusopalopem}, where the eclipsing binaries (including both EA and EW) and the RR Lyrae are shown in different colors. Principal components have been calculated on normalized light curves, so they depend only on the shape of the light curve and not on amplitude and period. Fig.~\ref{hukkusopalopem} thus shows that the light curves of RRc and eclipsing binaries differ in shape enough that the two groups appear as clusters in principal component space. Even though these clusters are not easily separated by eye, their degree of overlap is less than that witnessed in the Bailey diagram in Fig.~\ref{fig:bailey_diagram}. In principle one could build a classifier by placing a cut along one of the principal components shown in Fig.~\ref{hukkusopalopem}; this would result in a separating hyperplane in feature space. However, having access to the full feature space rather than to just the first three principal components, a classification algorithm is likely to perform substantially better than any fiducial cut we could place by hand. Moreover, as we will explain below, the algorithm we chose actually learns a separating hyperplane in feature space, which can be intuitively understood as an optimal cutoff on a linear combination of the features.

\subsection{(Regularized) Logistic Regression}
\noindent Logistic Regression is a widely used classification algorithm. Contrary to linear regression, there is no closed-form solution and one needs to solve it thanks to iterative convex optimization algorithms. Given a input  $X_i\in \mathbb{R}^d$, logistic regression classify it, i.e. returns a \REFEREE{boolean} label $Y_i\in\{0,1\}$, thanks to classification rule $f:\mathbb{R}^d\rightarrow \mathbb{R}$ such that $f(X_i)\geq 0\Rightarrow Y_i=1$ and $f(X_i)<0\rightarrow Y_i=0$.
In the linear case, $f$ is a linear operator of the form $f_{\beta}:x\rightarrow x^{T}\beta$.\vspace{0.2cm}\\ Intuitively, this algorithm seeks to identify the best linear separation for input data. However, this is achieved by a hyperplane that in general is not orthogonal with any of the coordinates of feature space, resulting in a model that has no non-zero coefficients. To obtain a smaller model by forcing some coefficients to zero we considered regularizing our classifier. This has the bonus of mitigating overfitting, even though this is not a major concern of ours at this stage; the main reason for inducing sparsity is to further promote interpretability.
There are several possible choices for the regularizer; however, when looking for a succinct description of the separating hyperplane, $L_1-$regularization performs better. Roughly speaking in fact it introduces a penalty upon the number of relevant features considered for the classification. 

\paragraph{Application}
\label{sec:analysis}
In our case, regularization allows us to limit the number of features included in the model, allowing us to understand which parts of a given light curve really matter for classification. We applied the logistic regression with the scikit-learn python package. 

\section{Results}\label{sec:res}
\noindent We trained our logistic regression model on a training set comprising $70\%$ of the CSS sample and evaluated its performance on a validation set containing the remaining $30\%$. The assignment to validation was stratified using \emph{StratifiedShuffleSplit} function from scikit-learn. We measured the performance according to a variety of metrics, as follows: \textit{accuracy}, ratio of stars correctly classified over the total; \textit{precision}, the fraction of stars classified as members of a given class who were actually members of that class; \textit{recall}, the fraction of actual members of a given class that were classified as such; and \textit{F-score}, the harmonic mean of precision and recall. We also obtained and plotted the precision-recall curve for our classifiers, which shows precision as a function of recall when the threshold used to convert the classifier's probabilistic prediction to a hard classification is varied. With the exception of accuracy, the metrics discussed above are class-dependent, i.e. they are in general different if one considers e.g. the fraction of actual RRc among the stars classified as RRc (precision for the RRc class) or the fraction of actual eclipsing binaries among the stars classified as eclipsing binaries (precision for the eclipsing binary class). A given decrease in precision for one class does not in general have the same cost as the same decrease for the other; polluting the sample of RRc on which distance measurements are based with eclipsing binaries is worse than the opposite. We thus report these metrics for each class.

\begin{figure}[H]
    \centering
    \begin{tabular}{cc}
    \includegraphics[width=0.4\linewidth]{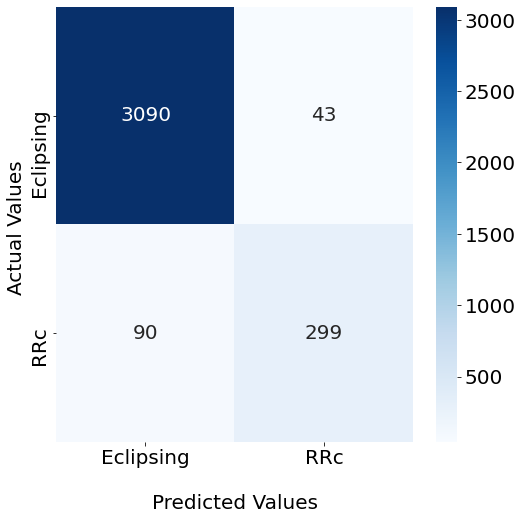} & \includegraphics[width=0.4\linewidth]{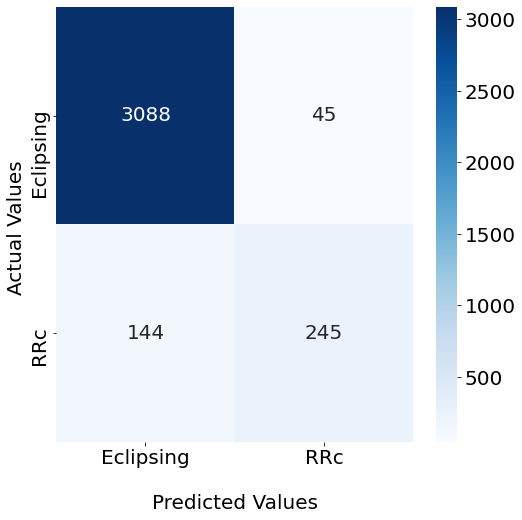}
    \end{tabular}
    \caption{Left: confusion matrix of our non-penalized classifier on the validation dataset (comprised of CSS light curves not seen in training). Right: confusion matrix of our L1 regularized classifier on the same validation dataset.}
    \label{fig:confu_nonpena}
\end{figure}

\begin{figure}[H]
\begin{tabular}{c}
    \centering
    \includegraphics[width=0.45\textwidth]{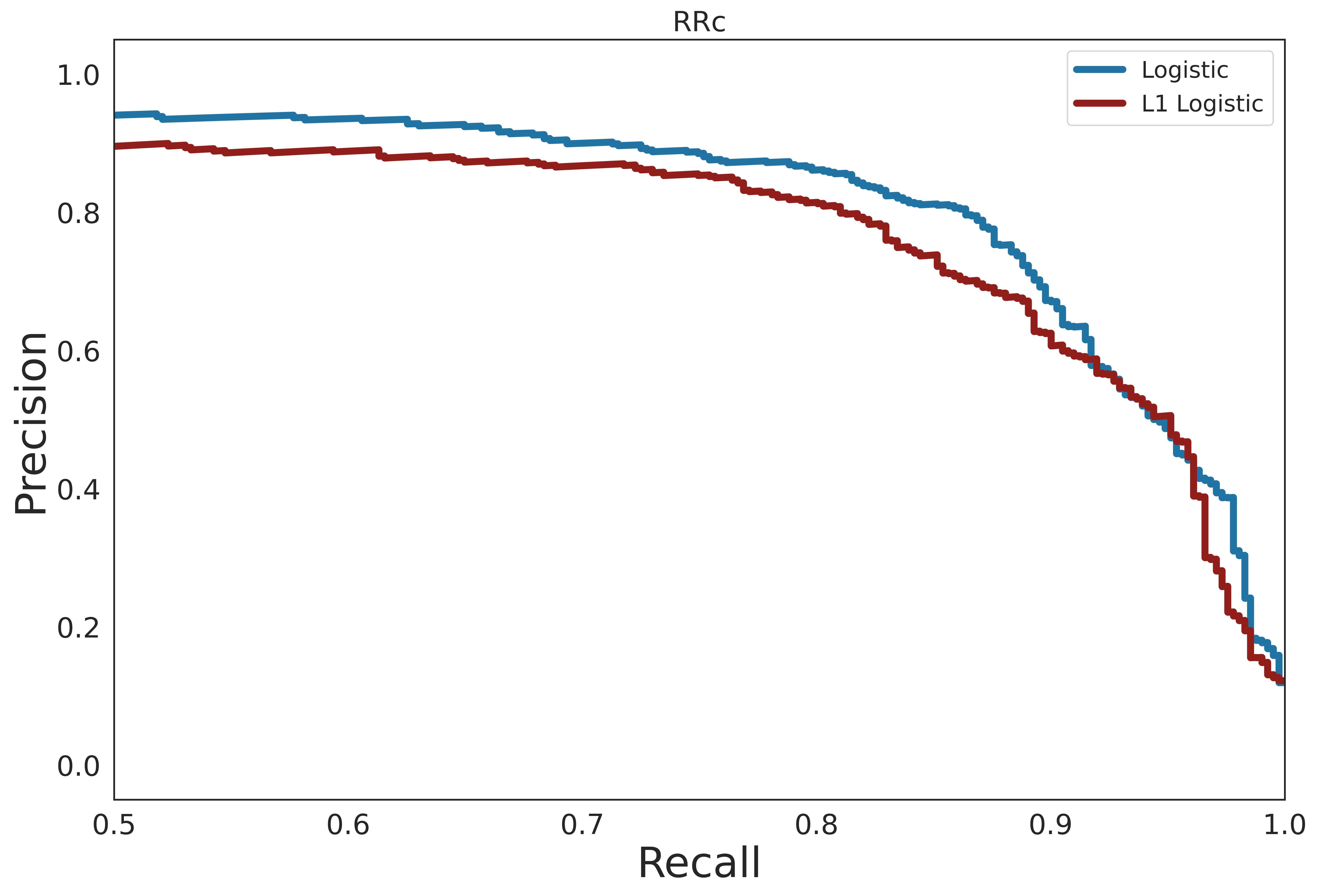}
    \includegraphics[width=0.45\textwidth]{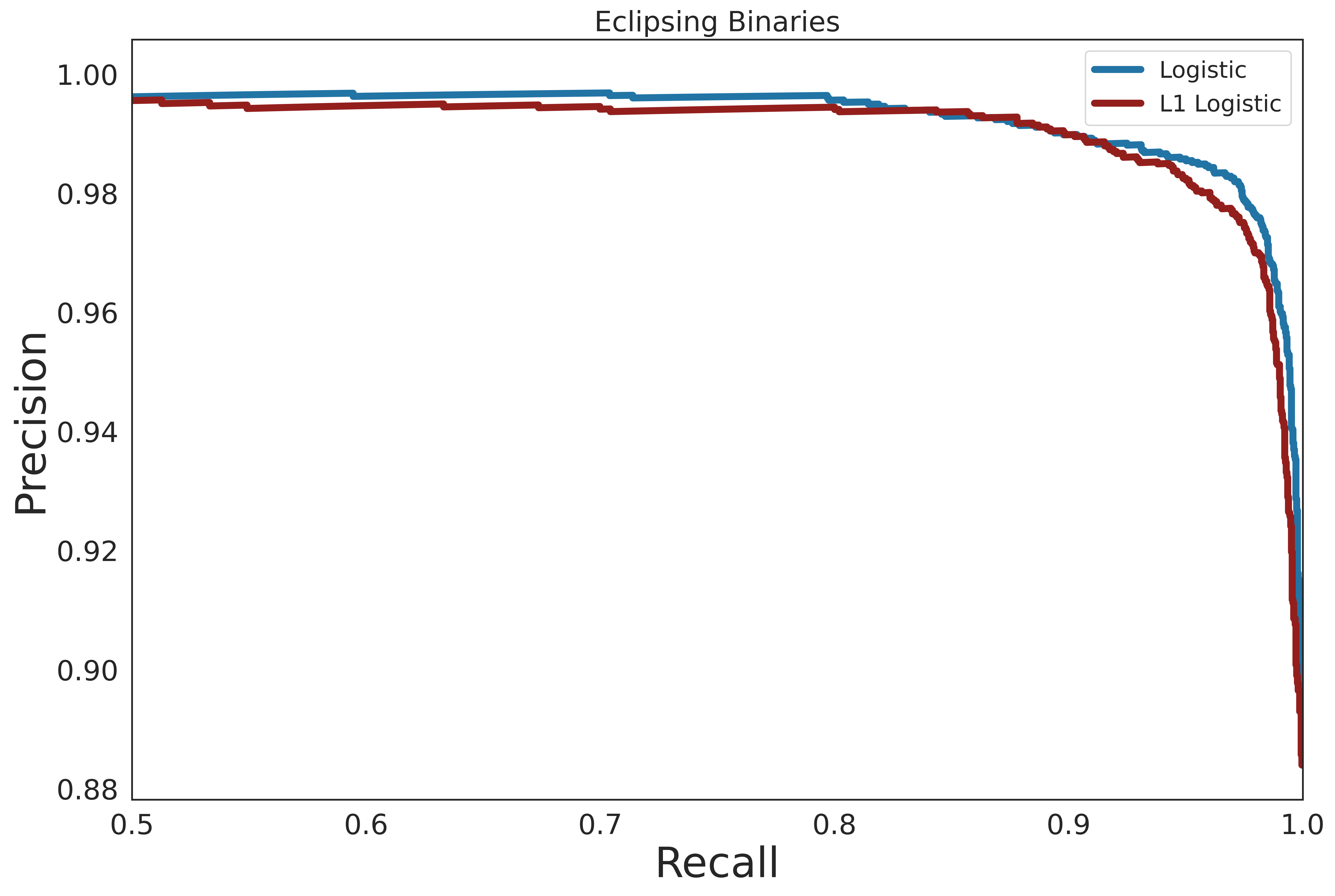}
\end{tabular}
    \caption{Precision-recall curve for our non-penalized model (blue) and for the L1 regularized version of our model (red) for the RRc and EB classes respectively on the left and right panel. Only the region with recall greater than $0.5$ is shown. The curve was calculated on the validation set (comprised of CSS light curves not seen in training). Note the difference on the y-axis between the two panels. } 
    \label{fig:recei_nonpena}
\end{figure}

\noindent The first model we present is a plain logistic regression, fit without a penalty term. The confusion matrix achieved on our test set is shown in Fig.~\ref{fig:confu_nonpena}. The performance metrics discussed above are reported in Tab.~\ref{tab:metrics} \REFEREE{for a standard threshold of 0.5}. A more detailed view of the classifier performance at various thresholds is available in Fig.~\ref{fig:recei_nonpena} through a precision-recall curve.

\begin{table}[H]
    \centering
    \caption{Performance metrics broken down by class (\REFEREE{first column}) of our non-penalized (\REFEREE{second column}) and L1 regularized model (\REFEREE{third column}) on the validation set.}
    \begin{tabular}{ccc}
        \hline 
        Metrics & $\text{Score}$ & $\text{Score}_{92}$\\
        \hline \hline
        Accuracy &  0.96 & 0.95\\
        \hline
        Recall (RRc) & 0.78 & 0.70\\
        Recall (EB) & 0.98 & 0.99\\
        \hline
        Precision (RRc) & 0.87 & 0.87\\
        Precision (EB) & 0.97 & 0.96\\
        \hline
        F-score (RRc) & 0.82 & 0.77\\
        F-score (EB) & 0.98 & 0.97\\
    \hline
    \end{tabular}
    
    \label{tab:metrics}
\end{table}

\noindent As discussed above, an advantage of logistic regression over other methods is the interpretability of the model. Compared to e.g. a deep neural network that can have millions of weights, our model is entirely specified by the values of the coefficients associated to each feature. Since each feature is a magnitude at a given phase we can visualize the coefficients as a function of the associated phase, interpreting them as weights on the light curve to be classified. This is shown in Fig.~\ref{fig:coeffi_nonpena}. A positive coefficient corresponds to an increased probability of being classified as an RR Lyrae when a given feature is increased, and vice-versa for a negative coefficient. This is similar to interpreting a classifier using a saliency map as in e.g. \cite{2021A&A...652A..19P} but with the enormous advantage that we do not need to compute the gradient locally for each instance, since our model is linear and the direction of the gradient is the same for every instance. In the absence of regularization, most coefficients differ from zero, but the coefficients are far from independent from each other, leading to contiguous regions of the light curve that are weighted similarly by the classifier. \REFEREE{In the bottom panel of Fig.~\ref{fig:coeffi_nonpena}} such regions appear.

\newpage
\section{Discussion}

\begin{figure}[H]
    \centering
    \begin{tabular}{c}
    \includegraphics[width=0.7\textwidth]{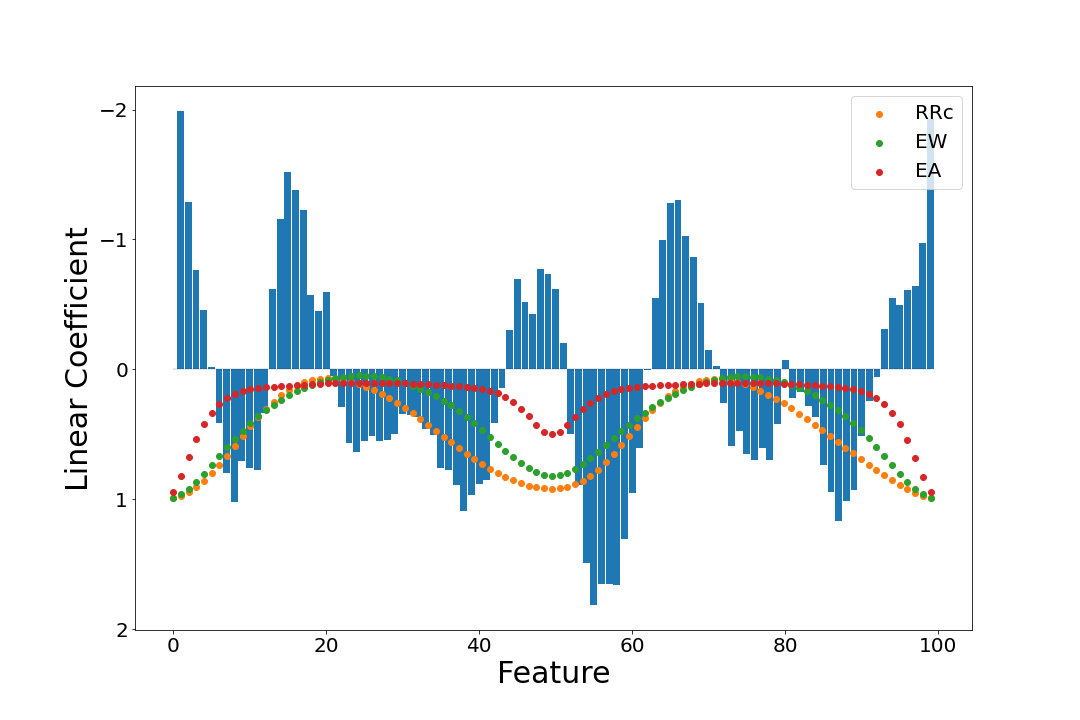} \\
    \includegraphics[width=0.7\textwidth]{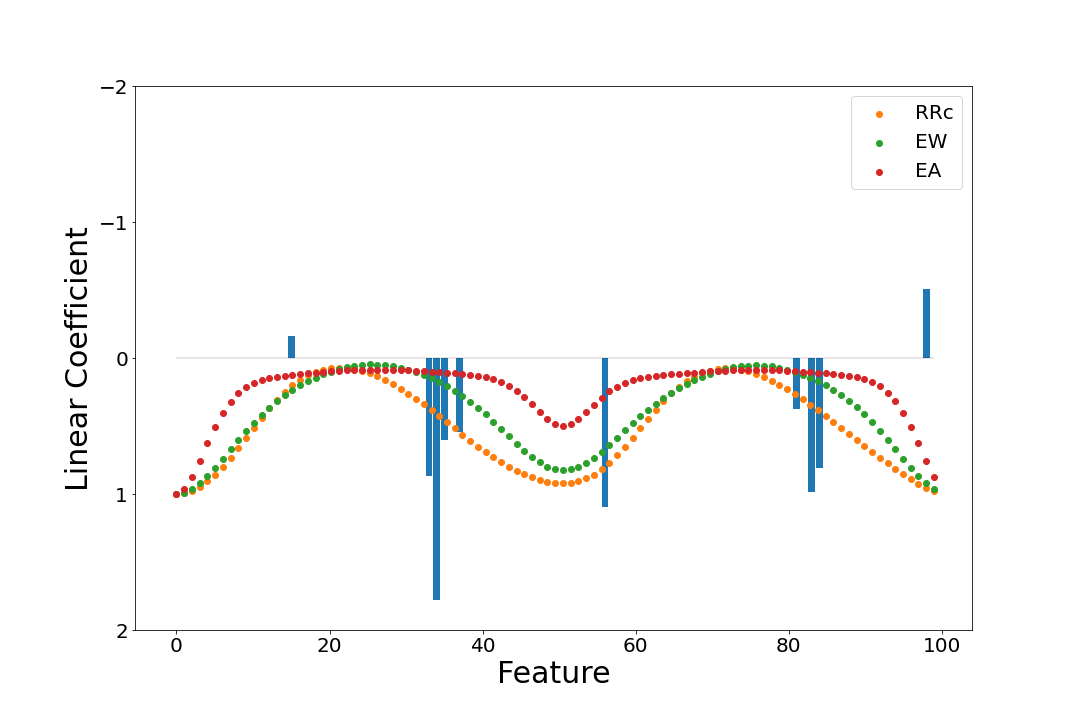}
    \end{tabular}
    \caption{Top: values of the linear coefficients of the logistic regression for the non-penalized model (blue bars). A positive coefficient means that an increased value of the magnitude at the corresponding phase would increase the confidence of classification as RRc. \REFEREE{Meanwhile, a negative coefficient increase the confidence of classification as EB}. Mean light curve of all RRc in the whole data set (orange), mean light curve of all EW (green) and EA (red) are shown as dots. Bottom: same for the L1 regularized model: the regularization strength was chosen so that $92\%$ of the coefficients became zero.}
    \label{fig:coeffi_nonpena}
\end{figure}

\noindent In Fig.~\ref{fig:coeffi_nonpena} we show the coefficients of our non-penalized logistic regression model (blue bars, top panel) together with the mean RRc light curve (orange points) and the mean eclipsing binary light curve (green points). The qualitative interpretation of the coefficients is as follows: increasing the value of the corresponding feature (moving one of the points in the light curve \REFEREE{downwards}) increases the probability of being classified as a RRc if the coefficient is positive, decreases it otherwise. Most coefficients are nonzero in the non-penalized model, making interpretation harder. The penalized model drastically reduces the number of non-zero coefficients, showing which parts of the curve are actually important for the classifier. These coincide roughly with the approach towards the minima of the light curve. This is showing that the classifier relies on the quicker rise (or fall) of the RRc light curve.

\REFEREE{\noindent The results obtained in the experimental analysis may seem at a first glance to be unsatisfactory when compared to results \REFEREE{of} other machine learning methods: in fact, as a baseline comparison we trained a random forest (with the default parameters of sklearn) obtaining a precision on the RRc of 0.91 and a recall of 0.96. However, it should be noted that there are trade-offs between the accuracy and interpretability of a model \citep{Interpretability(1),Interpretability(2),Interpretability(3)}. As there is no unambiguous definition for interpretability, there is no canonical quantification for this trade-off; however, important findings from the field of statistics confirm the existence of a statistical cost of interpretability \cite{Trade-off}. In our investigation, we strengthen the interpretability side, in fact we do not simply choose a classical statistical model such as regression, but increase its sparsity. Augmenting sparsity has an interest per se, for a survey around sparsity see \cite{Sparsity}.}

\begin{figure}
    \centering
    \includegraphics[width=0.7\linewidth]{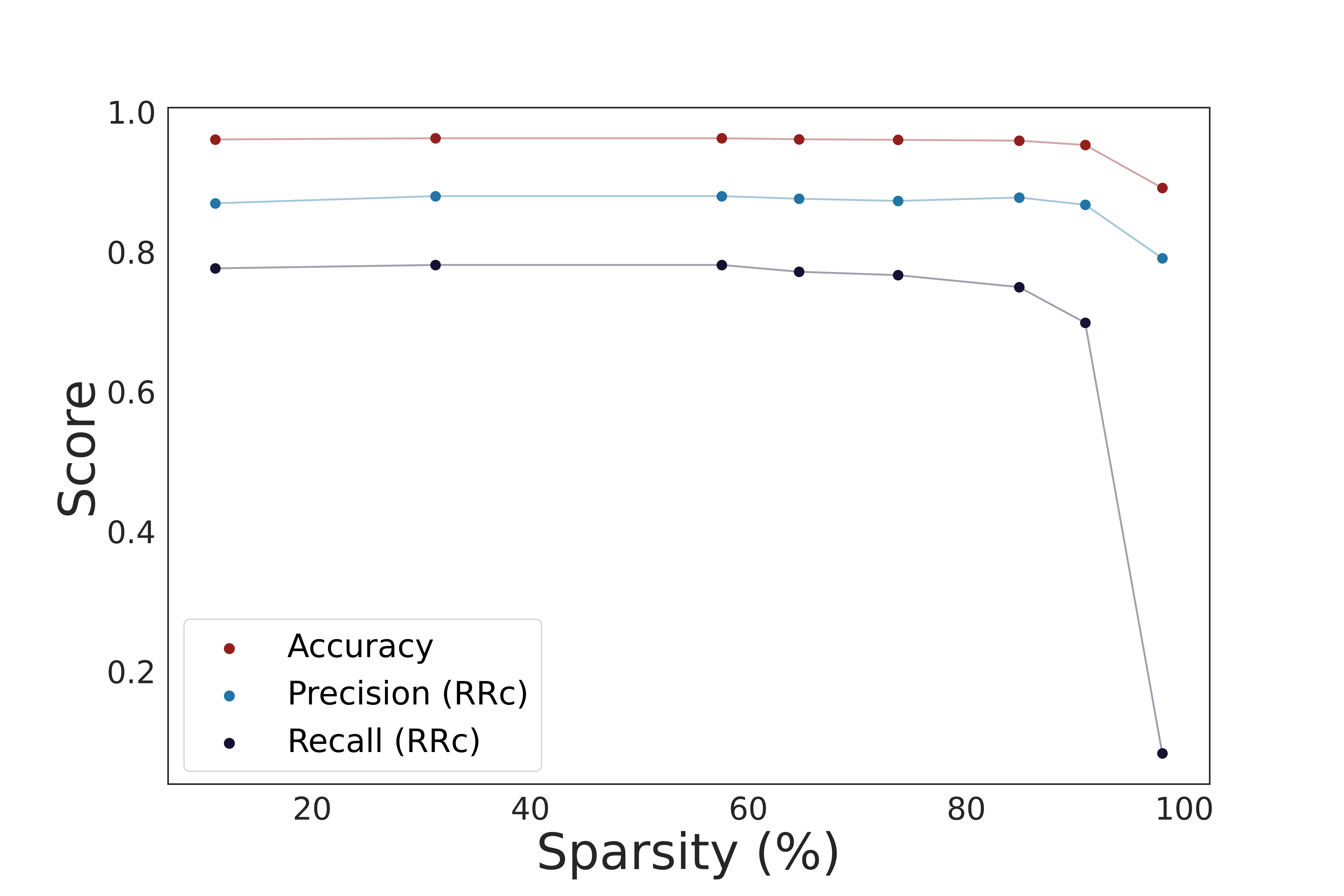}
    \caption{Performance metrics of our classifier (y axis) as a function of the fraction of coefficients that equal exactly zero (sparsity, x-axis). At over $90\%$ we reach a knee where the recall for the RRc class drops precipitously but for lower sparsity the three metrics are approximately constant.}
    \label{fig:penalty_metrics}
\end{figure}

\noindent Adding a sparsity-inducing penalty to the usual loss function optimized by logistic regression allows us to increase the number of coefficients that equal zero. This makes the classifier easier to understand and reduces the risk of overfitting. We chose to use a $L^1$ penalty, and we varied the weight assigned to the penalty term in the resulting loss function, leading to different regularization strengths. The effect on sparsity and on the performance of our classifier is shown in Fig.~\ref{fig:penalty_metrics}; this figure was used to select a regularization strength around the knee of the performance curve. Our choice resulted in a model with $92\%$ of the coefficients equal to zero. The relevant metrics are summarized in Tab.~\ref{tab:metrics}; regularization reduced our recall on the RRc class from $0.78$ to $0.70$ but did not impact precision.

\begin{figure}
\centering
\begin{tabular}{cc}
    
     \includegraphics[width=0.4\linewidth]{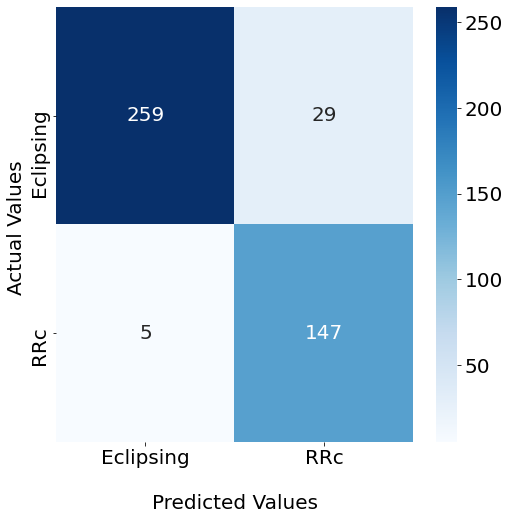}
     \includegraphics[width=0.4\linewidth]{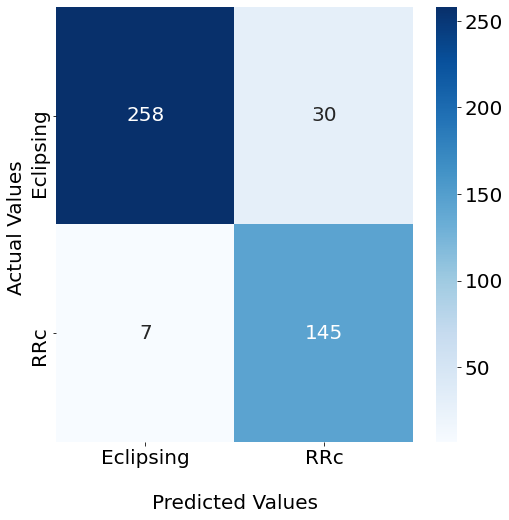}
    
    \end{tabular}
    \caption{Left: confusion matrix of our non-penalized classifier on the test dataset; right: confusion matrix of our L1 regularized classifier on the test dataset.}
    \label{fig:my_label}
\end{figure}

\vspace{0.2cm}\noindent In order to see how well our model predict on an unseen data from a different catalogue, \REFEREE{we tested both classifiers} on ASAS/ASAS-SN stars. The relevant metrics are summarized in Tab.~\ref{tab:metrics_penali_test} and the confusion matrices are shown in Fig.~\ref{fig:my_label}.

\begin{table}
    \centering
    \caption{\REFEREE{Performance metrics broken down by class (first column) of our non-penalized (second column) and L1 regularized model (third column) on the test set.}}
   \REFEREE{ \begin{tabular}{ccc}
        \hline
        Metrics & $\text{Score}$ & $\text{Score}_{92}$\\
        \hline \hline
        Accuracy &  0.92 & 0.92\\
        \hline
        Recall (RRc) & 0.96 & 0.95\\
        Recall (EB) & 0.91 & 0.97\\
        \hline
        Precision (RRc) & 0.85 & 0.83\\
        Precision (EB) & 0.98 & 0.90\\
        \hline
        F-score (RRc) & 0.90 & 0.89\\
        F-score (EB) & 0.94 & 0.93\\
        \hline
    \end{tabular}}
    \label{tab:metrics_penali_test}
\end{table}

\newpage
 

\section{\REFEREE{Summary and conclusions}}
\label{sec:conclusion}
\noindent Distance determination relying on RR Lyrae stars as standard candles is negatively affected by contamination due to eclipsing binaries. In this paper we have shown that a simple logistic regression classifier trained on CSS light curves can separate RRc variables from eclipsing binaries reaching $0.87$ precision at $0.78$ recall on a validation sample of unseen CSS light curves. On ASAS/ASAS-SN light curves recall becomes \REFEREE{$0.96$} and precision becomes  \REFEREE{$0.85$}, showing very good generalization even to data taken by a different instrument. It has to be noted that the fraction of RRcs in the ASAS/ASAS-SN sample is different from the CSS sample ($12\%$ VS \REFEREE{$35\%$}). Thus good generalization is due to our deliberate choice of a simple model which is unlikely to overfit the data.\vspace{0.2cm}\\
Since our features are normalized magnitudes at different phases, our classifier is making use only of the shape of the light curve, without receiving any information about its amplitude, absolute mean magnitude, or period. We are also able to visualize the logistic regression coefficients corresponding to the normalized magnitude at each phase point. Through our sparsification approach, we are able to identify how the classifier is using the light curve shape to cast its prediction. The features with non-zero coefficients correspond to the phase approaching the minima, suggesting that the steepness of the fall is used by the classifier. In terms of physical explanation, non-contact eclipsing binaries are in fact characterized by much flatter maxima (both the components are visible for an extended period of time) broken by steep minima (when one of the two components is occluded).\vspace{0.2cm}\\
\noindent We provide a precision-recall curve for the full classifier as well as for the penalized one, which in principle allows for a selection of the classification threshold based on the relative cost of false positives (eclipsing binaries classified as RR Lyrae) and false negatives (vice versa).\vspace{0.2cm}\\
We conclude with a few caveats and perspective improvements for the future. Our classifiers rely on turning the observed light curve, which is sampled at irregular intervals, into an equally spaced time series by means of interpolation. We carried out this interpolation based on Gaussian processes, which depend on several parameters to fully specify the covariance structure in terms of a kernel function. In the work described in this paper we fixed those parameters once and for all, even though this resulted in a handful of bad fits which were later discarded. There is room for improvement in this pre-processing of the data, especially if it is to be adapted to different data-sets.  In this content it is worth mentioning that current Zwicky
Transient Facility  and near
future Vera C. Rubin Observatory (\REFEREE{\citealt{masci2018zwicky}, \citealt{ivezic2019lsst}})
long-term variability surveys \REFEREE{(will) provide well sampled, long-baseline light curves in several optical- and near-infrared bands (see e.g. \citealt{marshall2017science})}. Thus providing the unique opportunity
to use physically rooted diagnostics (\REFEREE{amplitude} in different
photometric bands) to properly separate intrinsic from geometrical
variables.

Finally, our model should be compared to more expressive ones, such as a 1D convolutional neural network, which might attain better accuracy at the price of substantially increased model complexity.\vspace{0.3cm} 

\section*{Acknowledgments}
\REFEREE{\noindent This paper is supported by the  Fondazione  ICSC , Spoke 3 Astrophysics and Cosmos Observations, National Recovery and Resilience Plan (Piano Nazionale di Ripresa e Resilienza, PNRR) Project ID CN$00000013$ Italian
Research Center on High-Performance
Computing, Big Data and Quantum Computing funded by MUR Missione $4$
Componente $2$ Investimento $1.4$: Potenziamento strutture di ricerca e
creazione di campioni nazionali di R$\&$S (M4C2-19 ) - Next Generation EU
(NGEU).}
This project has received funding from the European Union's Horizon 
research and innovation program under the Marie Sk\l{}odowska-Curie grant agreement No. \REFEREE{$896248$}.
This research was made possible by a generous
donation from Eric and Wendy Schmidt by recommendation of the Schmidt Futures program

\vspace{0.4cm}\noindent\REFEREE{We would like to thank the anonymous referee for their insightful comments and suggestions, which greatly improved the content
and the readability of an early version of this manuscript.}

\newpage
\bibliography{main}{}
\bibliographystyle{aasjournal}

\end{document}